\pdfoutput=1
\RequirePackage{fix-cm}
\documentclass[smallextended]{svjour3}       

\smartqed  
\usepackage{mathptmx, cite,mathrsfs,bmpsize,tikz, relsize,color, natbib}      
\usepackage{graphicx, epsfig, float, rotating,latexsym,array,delarray,amsmath, amsfonts, subfig,ulem}
\usepackage[font=small,labelfont=bf,labelsep=space]{caption}

\usetikzlibrary{arrows}
\newcommand{\sI}{ \sum_{i=1}^m I_{M,i} }
\tikzset{
    line/.style={draw, -latex}
}
\begin{document}

\title{Intervention Strategies for Epidemics: Does Ignoring Time Delay Lead to Incorrect Predictions? \thanks{LBS and AB were supported by NSF DMS-1715651.}}
%

\titlerunning{Does Ignoring Time Delay Lead to Incorrect Predictions?}

\author{Adrienna Bingham \and Leah B.~Shaw}


\institute{A.~Bingham \at
              Department of Applied Science, William \& Mary, P.O.~Box 8795,  Williamsburg, VA 23187-8795 \\
              \email{anbingham@email.wm.edu}           
           \and
           L.~B.~Shaw \at
              Department of Mathematics, William \& Mary, P.O.~Box 8795,  Williamsburg, VA 23187-8795 \\
              \email{lbshaw@wm.edu}    
}

\date{Received: date / Accepted: date}

\maketitle

\begin{abstract}
Our paper investigates distributions of exposed and infectious time periods in an epidemic model and how applying a disease control strategy affects the model's accuracy. While ordinary differential equations are widely used for their simplicity, they incorporate an exponential distribution for time spent exposed or infectious. This allows for a high probability of unrealistically short exposed and infectious time periods. We propose that caution must be taken when applying intervention methods to basic models in order to avoid inaccurate predictions.  Delay differential equations, which use a delta distribution for exposed and infectious periods, can provide better realism but are more difficult to use and analyze. We introduce a multi-infected compartment model to interpolate between an ODE model with exponential distributions
and a DDE model with delta distributions in order to investigate the effect these distributions have on the dynamics of the system when an intervention method is also included. Using steady state stability and bifurcation analysis, this paper considers when simpler infectious disease models can be used versus when more realistic time periods must be incorporated. We find that the placement of control measures on subpopulations and the length of the time delay impacts the accuracy of the simpler models. 
\end{abstract}

\section{Introduction}
\label{intro}

Mathematical models have the potential to provide insight into the dynamics of an emerging disease. They can help determine the impact that control measures have before the epidemic becomes too severe \citep{anderson1992, brauer2001}. During the 2014 Ebola outbreak, a CDC modeling team immediately created two models (one with intervention and the other without) to determine how quickly control measures would slow the epidemic \citep{meltzer2014estimating, meltzer2016}. This helped them estimate the amount of time they had to gather and send resources. When predicting the course of an epidemic, modelers must balance simplicity and accuracy. It is important to incorporate key characteristics of the disease, such as transmission and intervention methods.
However, too many factors and complicated methods could make the system difficult and time consuming to solve. For example, a complicated mathematical model that incorporates more parameters would require more data to estimate those parameters. If the parameter values are not fit to data correctly, it reduces the accuracy of the model predictions. Furthermore, if there is a rapidly spreading epidemic, such as Ebola, time is of the essence. In \citet{meltzer2016}, it is stated that the turnaround time for answers to critical questions was less than one week. In addition, modelers must present their research to policy makers, and if the model is too complicated to communicate clearly, policy makers are less likely to accept the model and implement recommended controls in a timely manner, another problem CDC modelers faced during the 2014 outbreak \citep{meltzer2014estimating, meltzer2016}.

One of the simpler methods to model an epidemic is using ordinary differential equations (ODEs) such as the SEIR compartmental model, where a population is divided into compartments of individuals who are susceptible (those able to become infected), exposed (those who are infected but cannot spread the disease), infectious (those who are able to spread the disease), or recovered  (sometimes referred to as removed since they can no longer contract the disease) \citep{wearing2005, li2014, anderson1992, kermack1927}. Using ODEs involves assuming an exponential distribution for the time a person will spend in each compartment \citep{feng2007, sherborne2015, li2014, yang2008, huang2010, zhang2009}. However, ODEs can lead to underestimates of the infectious steady state due to the exponential distribution's large variance \citep{wearing2005, yang2008, feng2007, sherborne2015}. This occurs because a large variance can allow for a higher probability of an unrealistically short time spent, for example, exposed or infectious.

We would like to use a distribution with a much smaller variance for the time remaining in each compartment. For instance, the gamma and Poisson distributions have been used to model the exposed and infectious periods of various diseases such as smallpox, anthrax, and influenza due to the reduced size in variance of the distributions \citep{brookmeyer2005, eichner2003, nishiura2007}. This allows for more realistic times spent exposed or infectious. We focus on the delta distribution, which has constant exposed and infectious periods and a variance of zero. With this lack of variance, we limit the parameters we need to estimate. This type of distribution can be represented using delay differential equations (DDEs) \citep{huang2010, keeling2002}. However, it is more difficult to numerically evaluate DDEs than it is to evaluate ODEs because DDEs depend on a history of solutions to calculate the next time step. Also, most epidemic models are stiff systems due to processes occurring on multiple 
time scales \citep{shampine2001}. For instance, a model could include the average lifetime of an individual, which could span years, while also including the time span of a disease in days. Solving stiff systems becomes more difficult when time delays are involved. Therefore, it is our goal to highlight certain cases when it is best to use delay differential equations versus when the simpler ordinary differential equations can be used.

In this paper, we will compare an exponential model, which uses ordinary differential equations, and a delay differential equation model, with time delays in both an exposed and an infectious class, using a model with multiple exposed and infectious compartments to interpolate between the two. All three models will include quarantine of the infectious class as the main control measure. Section \ref{Mod} introduces the models, and Section \ref{SSBR} reports the steady states and basic reproduction numbers. In Section \ref{Underestimation}, we compare the effect of control in the exponential and delay models where we have used simulated data to fit the transmission rate for each model. As expected, we see an underestimation in the infectious steady state of the exponential system when applying a quarantine measure. Next, in Section \ref{CompartSS}, we explore whether the time delay in the exposed class or the time delay in the infectious class has the greatest impact on the infectious steady state. Then, in Section \ref{DelayLength}, we explore how the length of each time delay affects the difference between the exponential and the delay systems' infectious steady states. Finally, in Section \ref{DiffControls}, we compare our quarantine results with two other possible control measures, isolation of exposed individuals and vaccination of susceptibles.

\section{Models}
\label{Mod}
For this project we will use Ebola parameters from \citet{hu2015}. At the time of the 2014 outbreak, Ebola was spreading rapidly, so it was important that accurate models were produced quickly. Additionally, there were numerous intervention methods applied, including quarantine, hospitalization rates, and safe burial practices \citep{haas2014, pandey2014, meltzer2014estimating, meltzer2016}. Finally, because the incubation period and infectious period for Ebola are relatively long compared to diseases such as influenza and measles \citep{hethcote2000,wearing2005,safi2011}, we suspected the distributions of these periods could be especially important, although that will be explored in detail in Section \ref{DelayLength} .

In the following three models, we will focus on the intervention method of quarantine. A quarantine rate is applied to the infectious population, where a proportion of the infectious population will be isolated and unable to spread the disease. They will eventually enter the recovered class (see Fig.~\ref{fig:flow2}). Later, in Section \ref{DiffControls}, we will investigate what happens when control is applied to the exposed or susceptible classes.
\tikzstyle{int}=[draw, fill=blue!20, minimum size=1.5cm]
\begin{figure}[t]
\begin{center}
 \begin{tikzpicture}[node distance=3.25cm,auto,>=latex',scale=.75, every node/.style={scale=.75}][t]
    \node [int] (S) {\Huge \textbf{S}};
    \node (muin) [left of=S,node distance = 2 cm, coordinate] {$\mu$};
    \node [int] (E) [right of=S] { \Huge \textbf{E}};
    \node [int] (I) [right of=E] {\Huge \textbf{I}};
    \node [int] (R) [right of=I] {\Huge \textbf{R}};
    \node [int] (Q) [below of=I, node distance = 2.5 cm] {\Huge \textbf{Q}};

    \path[->] (muin) edge node {\Large $\mu$} (S);
    \path[->] (S) edge node {\Large $\beta$} (E);
    \path[->] (E) edge node {\Large $\sigma$} (I);
    \draw[->] (I) edge node {\Large $\gamma$} (R) ;
    \draw[->] (I) edge node {\Large $\alpha$} (Q) ;
    \draw[->] (Q) edge node {\Large $\gamma$} (R) ;

    \draw[->] (S.90) --++ (25:1 cm) node[midway]{\Large $\mu$};
    \draw[->] (E.90) --++ (25:1 cm) node[midway]{\Large $\mu$};
	\draw[->] (I.90) --++ (25:1 cm) node[midway]{\Large $\mu$};
    \draw[->] (R.90) --++ (25:1cm) node[midway]{\Large $\mu$};
    \draw[->] (Q.-90) --++ (-25:1.15 cm) node[pos = 0.7]{\Large $\mu$};

\end{tikzpicture} 
\caption{Flow chart for SEIQR model, where $\mu$  is the population birth and death rate, $\beta$ is the infection rate, $\sigma$ is the incubation rate, $\alpha$ is the quarantine rate, and $\gamma$ is the recovery rate }
\label{fig:flow2}
\end{center}
\end{figure}
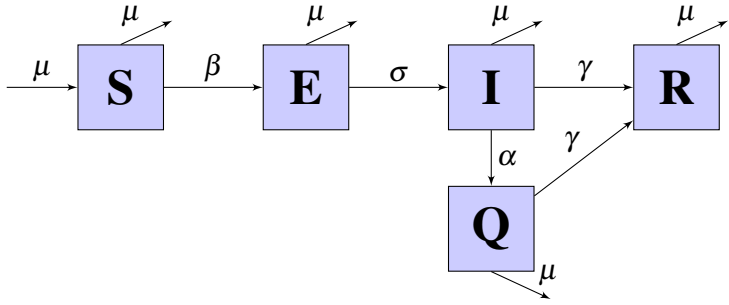

We first introduce the exponential version of the SEIR model with a quarantined class:
 \begin{equation}
\label{eq: exp}
\left \{ \begin{aligned}
\frac{dS}{dt} &= \mu-\beta SI-\mu S\\
\frac{dE}{dt} &= \beta SI-(\mu +\sigma) E\\
\frac{dI}{dt} &= \sigma E- (\mu +\alpha + \gamma)I\\
\frac{dQ}{dt} &= \alpha I-(\mu + \gamma) Q \\
\frac{dR}{dt} &= \gamma (I + Q)- \mu R
\end{aligned}\right.
\end{equation}
where $S$ is the proportion of susceptibles in the population, $E$ is the proportion of exposed, $I$ is the proportion of infectious, $Q$ is the proportion in quarantine, and $R$ is the proportion of removed individuals at time $t$. We have a transmission rate $\beta$, incubation rate $\sigma$ to advance from exposed to infectious, quarantine rate $\alpha$, recovery rate $\gamma$, and birth and death rate $\mu$. All rates are per day (see Table \ref{table:1}). For the quarantine rate, we use $\alpha = 0.05$. According to \citet{meltzer2016} and \citet{washington2015}, about 20\% of Ebola patients were put into Ebola treatment units between September and October of 2014. While other papers reference a higher quarantine rate for best case scenarios \citep{hu2015,hethcote2002}, we want to focus on a smaller $\alpha$ since our goal is not to guarantee epidemic extinction but rather to study the influence of control measures on accuracy in model predictions. 

\begin{table}[t]
\centering
\begin{tabular}{|c c c|} 
 \hline
 Parameter & Definition & Value\\ [0.5ex] 
 \hline\hline
 $\mu$ & Birth and death rate & 0.00005 days$^{-1}$ \\
 $\beta$ & Transmission rate & 0.278  days $^{-1}$\\ 
 $\sigma$ & Incubation rate & 0.10 days $^{-1}$ \\
 $\gamma$ & Recovery rate & 0.18 days $^{-1}$ \\
 $\alpha$ & Quarantine rate & 0.05 days$^{-1}$ \citep{meltzer2016} \\
 $\tau_1$ & Incubation delay $\left(\frac{1}{\sigma}\right)$ & 10 days \\ 
 $\tau_2$ & Recovery delay $\left(\frac{1}{\gamma}\right)$& 50/9 days \\
 $n $ & Number of exposed compartments & $1$ to $\infty$ \\
 $m$ & Number of infectious compartments & $1$ to $\infty$ \\
 \hline
\end{tabular}
\caption{Model parameters (derived from \cite{hu2015} except where indicated)}
\label{table:1}
\end{table}

Our delay differential equation model includes time delays $\tau_1$ and $\tau_2$, where $\tau_1$ represents the time spent in the exposed compartment and $\tau_2$ represents the time spent in the infectious compartment:
\begin{equation}
\label{eq: dde}
\left \{ \begin{aligned}
\frac{dS}{dt} = \; & \mu-\beta SI-\mu S\\
\frac{dE}{dt} = \; & \beta SI - \beta S(t-\tau_1)I(t-\tau_1)e^{-\mu \tau_1}-\mu E\\
\frac{dI}{dt} = \; & \beta S(t-\tau_1)I(t-\tau_1)e^{-\mu\tau_1}   \\
 & - \beta S(t-\tau_1 - \tau_2)I(t-\tau_1-\tau_2)e^{-\mu(\tau_1+\tau_2)-\alpha\tau_2} - (\mu+\alpha) I \\
\frac{dQ}{dt} = \; & \alpha I-\mu Q - \beta S(t-\tau_1 - \tau_2)I(t-\tau_1-\tau_2)e^{-\mu(\tau_1+\tau_2)}(1-e^{-\alpha\tau_2})\\
\frac{dR}{dt} = \; & \beta S(t-\tau_1 - \tau_2)I(t-\tau_1-\tau_2)e^{-\mu(\tau_1+\tau_2)}-\mu R\\
\end{aligned} \right . .
\end{equation}
The rate of entry into the infectious class from the exposed class at time $t$ is $\beta S(t-\tau_1)I(t-\tau_1)e^{-\mu\tau_1}$, where a person who becomes infectious at $t$ was exposed at time $t-\tau_1$ and remained exposed for length of time $\tau_1$ \citep{cooke1999}. The factor $e^{-\mu\tau_1}$ is the probability an exposed individual will survive to reach the infectious class \citep{beretta2011,kaddar2011,li2014}. 
Similarly, entry into the recovered class from the infectious class occurs at rate
\begin{equation*}
\beta S(t-\tau_1 - \tau_2)I(t-\tau_1-\tau_2)e^{-\mu(\tau_1+\tau_2)}e^{-\alpha\tau_2},
\end{equation*}
where a person recovering at time $t$ was exposed at time $t-\tau_1-\tau_2$, and the probability the person survives to reach the recovered class is $e^{-\mu(\tau_1+\tau_2)}$. The probability a person does not become quarantined and moves straight from infectious to the recovered class is $e^{-\alpha\tau_2}$. In the equation for $Q$, $1-e^{-\alpha\tau_2}$ represents the probability an individual did become quarantined. 

Note that although it is possible to exclude $\frac{dE}{dt}$ and $\frac{dR}{dt}$ because they do not affect the other compartments, we include these equations for completeness and to emphasize the fact that we have a closed system.  In other words, our total population is always 100\%.

To be able to compare the delay model to the exponential model, the time delays must correspond to the average times in each compartment in the exponential model. The transition rate from the exposed compartment to the infectious compartment is $\sigma$, so the average time spent in the exposed class is $\frac{1}{\sigma}$. This is our $\tau_1$.  Similarly, $\tau_2 = \frac{1}{\gamma}$.

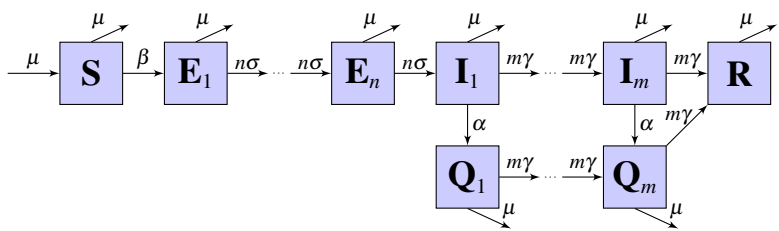
\begin{figure}[t]
\begin{center}
 \begin{tikzpicture}[node distance=2.5cm,auto,>=latex',scale=.55, every node/.style={scale=.55}]
    \node [int] (S) {\Huge \textbf{S}};
    \node (muin) [left of=S,node distance = 2. cm, coordinate] {$\mu$};
    \node [int] (E1) [right of=S] { \Huge $\textnormal{\textbf{E}}_{\scriptscriptstyle 1}$};
    \node [fill=white, minimum size=.5cm] (dots1) [right of =E1,node distance = 2 cm] {$\cdots$};
    \node [int] (En) [right of=dots1,node distance = 2 cm] { \Huge $\textnormal{\textbf{E}}_{\scriptscriptstyle n}$};
    \node [int] (I1) [right of=En] { \Huge $\textnormal{\textbf{I}}_{\scriptscriptstyle 1}$};
    \node [fill=white, minimum size=.5cm] (dots2) [right of =I1,node distance = 2 cm] {$\cdots$};
    \node [int] (Im) [right of=dots2,node distance = 2 cm] { \Huge $\textnormal{\textbf{I}}_{\scriptscriptstyle m}$};
    \node [int] (R) [right of=Im] {\Huge \textbf{R}};
    \node [int] (Q1) [below of=I1, node distance = 2.5 cm] { \Huge $\textnormal{\textbf{Q}}_{\scriptscriptstyle 1}$};
    \node [fill=white, minimum size=.5cm] (dots3) [right of =Q1,node distance = 2 cm] {$\cdots$};
    \node [int] (Qm) [below of=Im, node distance = 2.5 cm] { \Huge $\textnormal{\textbf{Q}}_{\scriptscriptstyle m}$};

    \path[->] (muin) edge node {\Large $\mu$} (S);
    \path[->] (S) edge node {\Large $\beta$} (E1);
    \path[->] (E1) edge node {\Large $n\sigma$} (dots1);
    \path[->] (dots1) edge node {\Large $n\sigma$} (En);
    \draw[->] (En) edge node {\Large $n\sigma$} (I1) ;
    \draw[->] (I1) edge node {\Large $m\gamma$} (dots2) ;
    \draw[->] (dots2) edge node {\Large $m\gamma$} (Im) ;
    \draw[->] (Im) edge node {\Large $m\gamma$} (R) ;
    \draw[->] (I1) edge node {\Large $\alpha$} (Q1);
    \draw[->] (Q1) edge node {\Large $m\gamma$} (dots3) ;
    \draw[->] (dots3) edge node {\Large $m\gamma$} (Qm) ;
    \draw[->] (Im) edge node {\Large $\alpha$} (Qm) ;
    \draw[->] (Qm) edge node[left, pos=0.7] {\Large $m\gamma$} (R) ;

    \draw[->] (S.90) --++ (25:1cm) node[midway]{\Large $\mu$};
    \draw[->] (E1.90) --++ (25:1cm) node[midway]{\Large $\mu$};
    \draw[->] (En.90) --++ (25:1 cm) node[midway]{\Large $\mu$};
	\draw[->] (I1.90) --++ (25:1 cm) node[midway]{\Large $\mu$};
    \draw[->] (Im.90) --++ (25:1 cm) node[midway]{\Large $\mu$};
    \draw[->] (R.90) --++ (25:1 cm) node[midway]{\Large $\mu$};
    \draw[->] (Q1.-90) --++ (-25:1.15 cm) node[pos = 0.7]{\Large $\mu$};
    \draw[->] (Qm.-90) --++ (-25:1.15 cm) node[pos = 0.7]{\Large $\mu$};
    
\end{tikzpicture} 
\end{center}
\caption{Flow chart for multi-infected compartment model, where $n$ is the number of exposed compartments and $m$ is the number of infectious/quarantine compartments. \label{fig:compflow}}
\end{figure}

Our third model, the multi-infected compartment model, breaks the disease classes of the exponential model into $n$ number of exposed compartments, $m$ number of infectious compartments, and $m$ number of quarantine compartments. This method is similar to the method used in \citet{wearing2005}. However, their model focused on contact tracing and divided the infectious individuals into asymptomatic and symptomatic categories while putting individuals who were exposed, whether they obtained the disease or not, into quarantine. In our model, the number of infectious and quarantine compartments will always be equal (see Fig.~\ref{fig:compflow}) since recovery in the infectious class will take just as long in the quarantine class: 
\begin{equation}
\label{eq: comp}
\left \{ \begin{aligned}
\frac{dS}{dt} &= \mu-\beta S\left(\sum_{i = 1}^m I_i\right)-\mu S\\
\frac{dE_1}{dt} &= \beta S\left(\sum_{i = 1}^m I_i\right)-\mu E_1-n\sigma E_1\\
\vdots \\
\frac{dE_n}{dt} &= n\sigma E_{n-1}-\mu E_n-n\sigma E_n\\
\frac{dI_1}{dt} &= n\sigma E_n-\mu I_1-(\alpha + m\gamma)I_1\\
\vdots \\
\frac{dI_m}{dt} &= m\gamma I_{m-1} - (\mu + \alpha +m\gamma) I_m \\
\frac{dQ_1}{dt} &= \alpha I_1 - (\mu+\gamma m)Q_1 \\
\vdots \\
\frac{dQ_m}{dt} &= \alpha I_m + \gamma m Q_{m-1} - (\mu + \gamma m)Q_m \\
\frac{dR}{dt} &= \gamma m (I_m+Q_m)-\mu R\\
\end{aligned} \right. .
\end{equation}
The term $\sum_{i = 1}^m I_i$ represents the total number of infectious individuals in the population at time $t$. Again, we need to make sure the average time spent in each class is equivalent to the other models. If there are $n$ compartments in the exposed class and the average time to become infectious is $\frac{1}{\sigma}$, then the average time in each exposed compartment is $\frac{1}{n\sigma}$, making the rate to progress out of each compartment $n\sigma$. Similarly, the rate to progress from each infectious compartment is $m\gamma$.

\begin{figure}[t]
\begin{center}
\includegraphics[scale=.50]{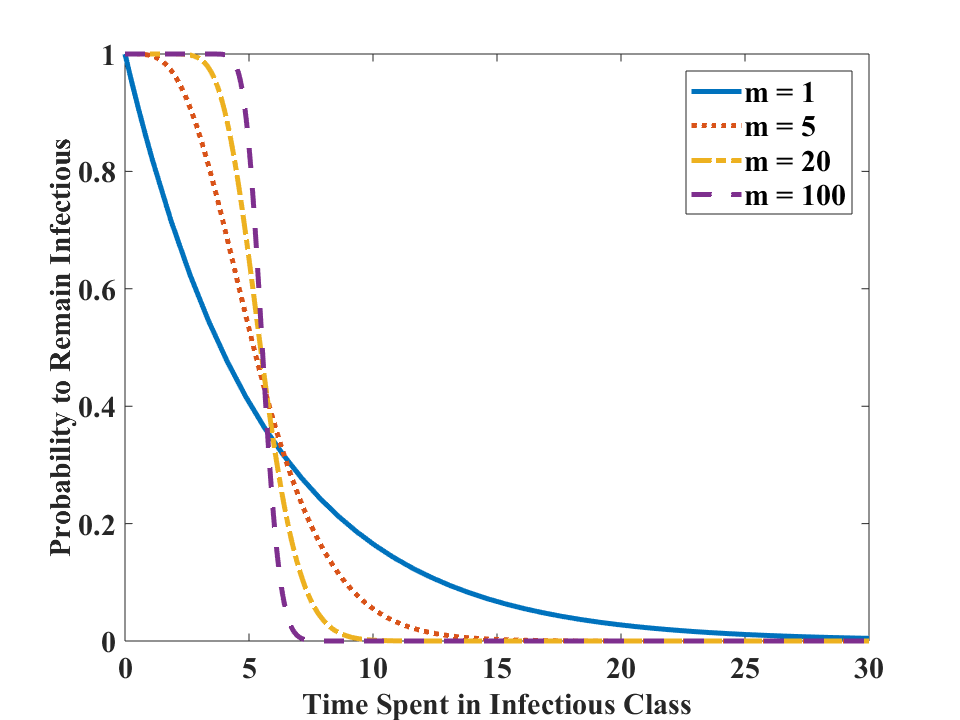} \\
\end{center}
\caption{Interpolation between the exponential and the delay model by the multi-infected compartment model. When $m$, the number of infectious compartments, is 1, the model assumes an exponential distribution. As $m$ increases, the model approaches a constant infectious period, which is used by the delay model
\label{fig:change}}
\end{figure}

This model uses a gamma distribution for time spent exposed and infectious, with the number of compartments affecting the shape of the distribution. As $n$ and $m$ increase towards infinity, the model approaches a delta distribution for time spent exposed and infectious, allowing us to observe the effect of constant exposed and infectious periods (see Fig.~\ref{fig:change}) \citep{sherborne2015, leemis2011, cox1965, feng2007}. When $n=m=1$, the model uses an exponential distribution for time spent exposed and infectious, equivalent to Eq.~(\ref{eq: exp}). Therefore, we will use this model to gradually change from the exponential model to the delay model and investigate which delays are causing the greatest change in dynamics.

\section{Steady States and Basic Reproduction Numbers}
\label{SSBR}
As mentioned before, some models that include multiple time spans can be stiff. While there are solvers made to adjust the time steps with respect to the time delays \citep{shampine2001}, our system proved to be too stiff for numerically accurate simulations. Our system also includes a combination of multiple time delays, increasing the difficulty in obtaining numerics. We focus primarily on analytics and find the endemic steady state of each model, with emphasis on the number of infectious individuals. We then calculate the basic reproduction number for each model. 
\subsection{Exponential Model}
The exponential system yields a disease free state of $(S_E^0, E_E^0, I_E^0, Q_E^0, R_E^0)=$ \newline $(1, 0, 0, 0, 0)$. There is also an endemic steady state $(S_E^*, E_E^*, I_E^*, Q_E^*, R_E^*)$, where our main focus will be on the infectious steady state (see Appendix \ref{ExpSS} for full steady state), 
\begin{equation}
\label{eq:Iess}
 I_E^* =\frac{\mu}{\alpha+\mu} \left(\frac{\sigma}{\sigma+\mu}\right)\left[1-\left(\frac{\gamma}{\gamma +\alpha+ \mu}\right)\right] -\frac{\mu}{\beta}.
\end{equation}

We can find the basic reproduction number using the method outlined by \citet{van2002}, obtaining
\begin{equation*}
R_0 = \frac{\beta\sigma}{(\sigma+\mu)(\alpha+\gamma+\mu)}.
\end{equation*}
For the Ebola parameters listed in Table \ref{table:1}, $R_0 > 1$. This implies that the disease free state is unstable under these parameters. By numerically calculating the Jacobian at the endemic equilibrium and its eigenvalues, it can be shown that the endemic steady state is stable under these parameters.

\subsection{Delay Differential Equation Model}
The disease free steady state of the delay differential equation model is identical to the exponential system, and the endemic steady state (see Appendix \ref{DDESS} for the full steady state) includes the infectious steady state  
\begin{equation}
\label{eq:Idss}
I_D^* =\frac{\mu}{\alpha+\mu} e^{-\mu\tau_1}\left[1-e^{-(\mu+\alpha)\tau_2}\right] -\frac{\mu}{\beta}.
\end{equation}

In order to find the basic reproduction number, we take the limit of the basic reproduction number of the multi-infected compartment model (see next subsection) as $n$ and $m$ approach infinity \citep{wearing2005}. Also, recall that $\tau_1 = 1/\sigma$ and $\tau_2 = 1/\gamma$. This results in 
\begin{equation}
R_0 =\frac{\beta}{\mu+\alpha}\left[e^{-\mu\tau_1}\right]\left[1-e^{-(\mu+\alpha)\tau_2}\right].  
\end{equation}

\subsection{Multi-Infected Compartment Model}

The disease free steady state of the multi-infected compartment model is similar to the previous two models: $(S_M^0, E_{M,1}^0, E_{M,2}^0,...,$
$ E_{M,n}^0, I_{M,1}^0, I_{M,2}^0$,...,$I_{M,m}^0, Q_{M,1}^0, Q_{M,2}^0,...,$ \newline $Q_{M,m}^0,R_M^0)
 = (1,0,0,...,0,0,0,...,0,0,0,...,0,0)$.

We now outline how we find the endemic steady state. Our main focus will be on the sum of all infectious compartments at steady state, $\sum_{i=1}^m I_{M,i}^*$, so we will express other quantities in terms of that sum.
First, we note that at steady state, 
\begin{equation*}
S_M^* = \frac{\mu }{\beta \sum_{i=1}^m I_{M,i}^* + \mu}
\end{equation*}
and 
\begin{equation}
\label{eq:E1}
E_{M,1}^* = \frac{\beta \mu \sum_{i=1}^m I_{M,i}^*}{(\mu + n \sigma)(\beta \sum_{i=1}^m I_{M,i}^* + \mu)}.
\end{equation}

Now, we find the steady state relationships between exposed classes $E_{M,1}^*, E_{M,2}^*,$\newline $...E_{M,n}^*$, where
 \begin{equation*}
 E_{M,i}^* = \left(\frac{n\sigma}{\mu + n \sigma}\right)^{i-1} E_{M,1}^*
 \end{equation*}
for $i\in\left\{2,\ldots,n\right\}$, indicating that 
 \begin{equation*}
\sum_{i=1}^n E_{M,i}^* = E_{M,1}^* \sum_{j=0}^{n-1} \left(\frac{n\sigma}{\mu + n \sigma}\right)^{j}.
 \end{equation*}
This a finite geometric series, and after substituting $E_{M,1}^*$ with (\ref{eq:E1}), we find that
\[
\sum_{i=1}^n E_{M,i}^* = \frac{\beta \sum_{i=1}^m I_{M,i}^* \left[1-\left(\frac{n\sigma}{n\sigma+\mu}\right)^n\right]}{\beta \sum_{i=1}^m I_{M,i}^* + \mu}.
\]

We now consider the steady state for the quarantined and recovered population $\sum_{i=1}^m Q_i+ R$.  In general,
\begin{equation*}
\begin{aligned}
\frac{d}{dt}\left[\sum_{i=1}^m Q_i+ R\right] &= \alpha  \sum_{i=1}^m I_i - \mu \sum_{i=1}^m Q_i -\mu R + m\gamma I_m.
\end{aligned}
\end{equation*}
At steady state
 \begin{equation*}
I_{M,i}^* = \left(\frac{m\gamma }{\mu+\alpha+m\gamma}\right)^{i-1} I_{M,1}^*
\end{equation*}
for $i\in\left\{2,\ldots,n\right\}$ and where 
 \begin{equation*}
I_{M,1}^* = \frac{n\sigma E_{M,n}^*}{\mu+\alpha+m\gamma}.
\end{equation*}
Therefore, 
\begin{equation*}
\left(\sum_{i=1}^m Q_i+ R\right)_M^* = \frac{\alpha }{\mu}\sum_{i=1}^m I_{M,i}^*+ \left(\frac{m\gamma}{\mu+\alpha+m\gamma}\right)^m \left(\frac{n\sigma}{\mu+n\sigma}\right)^n\left(\frac{\beta \sum_{i=1}^m I_{M,i}^*}{\beta \sum_{i=1}^m I_{M,i}^* + \mu }\right).
\end{equation*}

Finally, using the property of a closed system, $S_M^* + \sum_{i=1}^n E_{M,i}^* + \sum_{i=1}^m I_{M,i}^* +$\newline $\left(\sum_{i=1}^m Q_i+ R\right)_M^* = 1$, we solve for $\sum_{i=1}^m I_{M,i}^*$:
\begin{equation}
\label{eq:Iss}
\sI ^*  =\frac{\mu}{\alpha+\mu} C_E C_I -\frac{\mu}{\beta},
\end{equation}
where we define constants
\begin{equation}
\label{eq:CE}
C_E  =\left(\frac{n\sigma}{n\sigma+\mu}\right)^n
\end{equation}
and 
\begin{equation}
\label{eq:CI}
C_I  = 1-\left(\frac{m\gamma}{m\gamma +\alpha+ \mu}\right)^m.
\end{equation}

Again using the method from \citet{van2002} (see Appendix \ref{R0MIC} for details), the $R_0$ value is found to be 
\begin{equation}
R_0 =  \frac{\beta}{\mu+\alpha}C_EC_I.
\end{equation}

\section{Example of Underestimation of Infectious Population in Exponential System} 
\label{Underestimation}

When modeling a new epidemic, one of the more complicated parameters to estimate is the transmission rate \citep{hu2015, hethcote2000}. It is easier to estimate from data how many people are infected, how many people are being quarantined, or how many people have recovered. It is not as simple to calculate how often infectious people come into contact with susceptible people, and even more complicated to determine if the disease was transmitted with that contact.  To mirror the process of estimating the transmission rate of a real disease using a simple model before applying a control, we will fit our transmission rate $\beta$ in the exponential model to simulated data coming from the infectious steady state of the delay system.  We will then investigate whether applying a control leads to an underestimation in the infectious steady state of the exponential system like that observed by \citet{wearing2005}.

\begin{figure}[t]
\subfloat[]{\label{fig:scheme}\includegraphics[scale=.22]{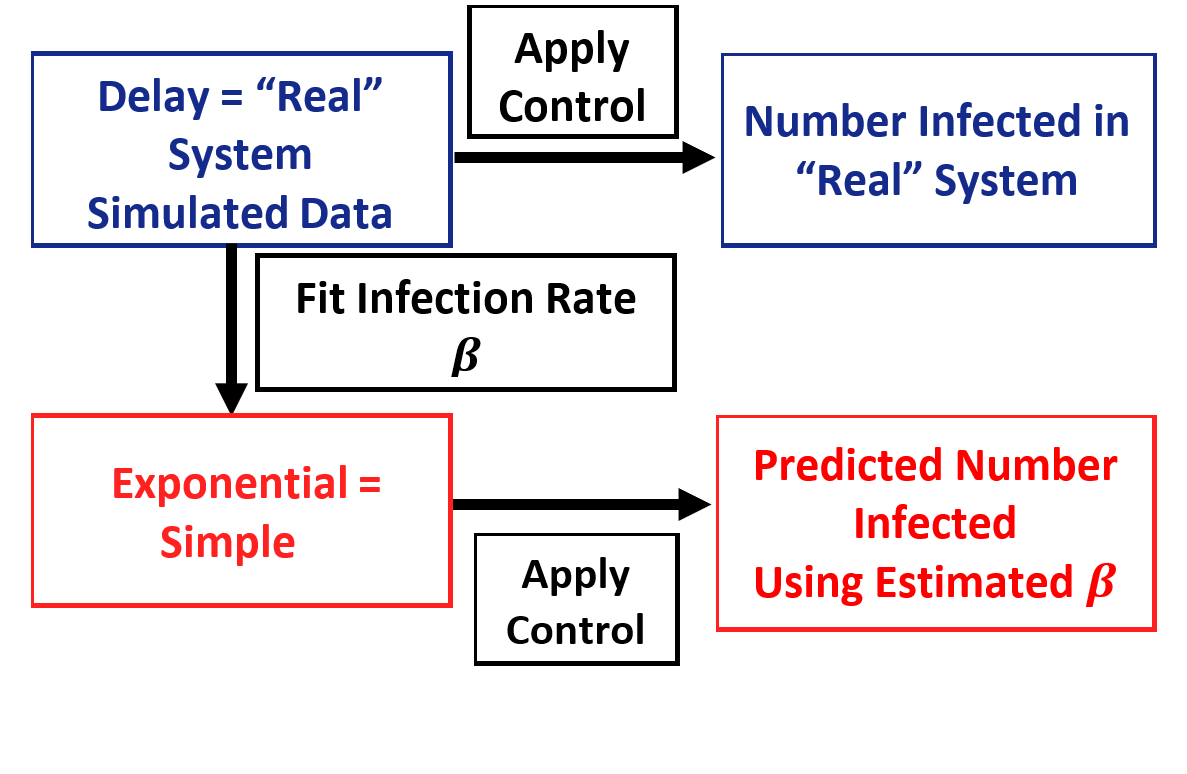}}
\subfloat[]{\label{fig:BetaEst}\includegraphics[scale=.27]{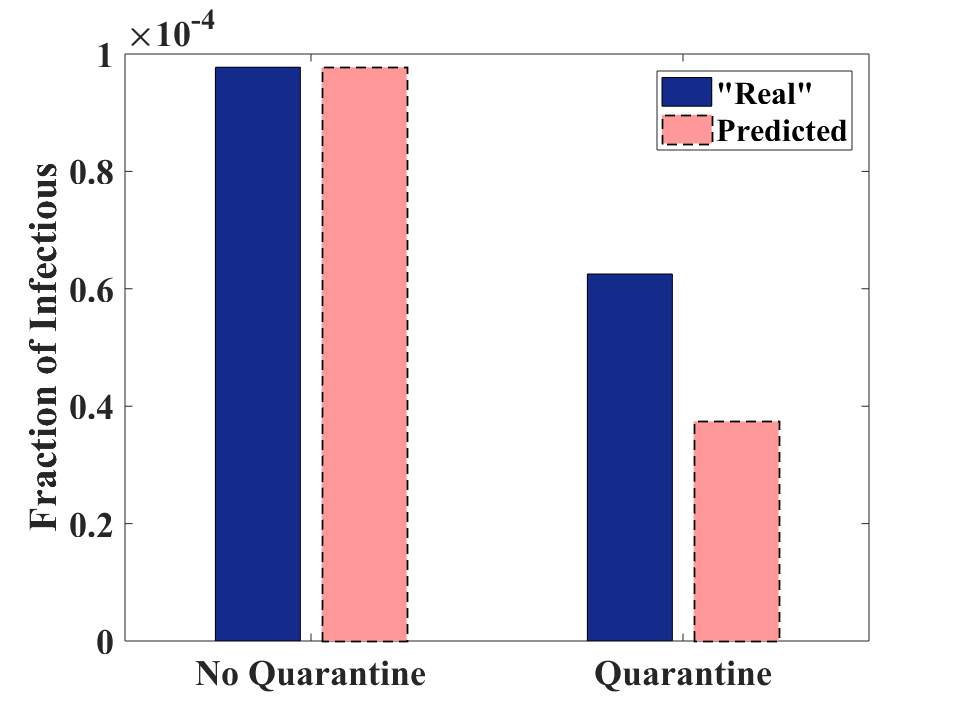}}
\caption{(a) Schematic for fitting $\beta$ and applying control to measure underestimation in exponential system. (b) Numerical results for fitting $\beta$, applying quarantine control, and comparing the ``real" delay system and the ``predicted" exponential system}
\label{fig:UnderEst}
\end{figure}

The procedure is shown in Fig.~\ref{fig:scheme}.
 We first begin by matching the infectious steady states of the exponential system (\ref{eq:Iess}) and the delay system (\ref{eq:Idss}) without any control ($\alpha = 0$). We will take our exponential system to be the ``predicted" or ``simple" system and the delay system to be our ``real" system. This is because the delay system is likely more accurate with its delta distribution and correspondingly small variance in exposed and infectious periods. Therefore, if $I^*_{E} = I^*_{D}$, then 

\begin{equation*}
\frac{\mu}{\alpha+\mu} \left(\frac{\sigma}{\sigma+\mu}\right)\left[1-\left(\frac{\gamma}{\gamma +\alpha+ \mu}\right)\right] -\frac{\mu}{\beta _{\textnormal{est}}} = \frac{\mu}{\alpha+\mu} e^{-\mu\tau_1}\left[1-e^{-(\mu+\alpha)\tau_2}\right] -\frac{\mu}{\beta_{\textnormal{real}}},
\end{equation*}
where $\beta_{\textnormal{real}}$ is the transmission rate in the delay system and $\beta_{\textnormal{est}}$ is the estimated transmission rate that allows the exponential system to match the delay system.
Solving, we find
\begin{equation}
\label{eq:Best}
 \beta_{\textnormal{est}} =  \frac{(\mu+\sigma)(\gamma + \mu)\beta_{\textnormal{real}}\mu}{(\sigma + u)(\gamma + \mu)(\beta_{\textnormal{real}} e^{-(\mu(\tau_1+\tau_2))}-\beta_{\textnormal{real}} e^{-\mu\tau_1}+\mu)+\sigma \beta_{\textnormal{real}} \mu}.
\end{equation}

We then substitute this estimated $\beta$ in for the transmission rate of our exponential steady state, (\ref{eq:Iess}), with $\alpha \neq 0$: 
\begin{equation}
\label{eq:BestSS}
I^*_{\textnormal{est}} = \frac{-\mu\sigma\alpha}{(\mu+\sigma)(\gamma+\mu)(\alpha+\gamma+\mu)}+e^{-\mu\tau_1}(1-e^{-\mu\tau_2})-\frac{\mu}{\beta_{\textnormal{real}}}
\end{equation}
If we substitute our parameter values from Table 1 into (\ref{eq:Idss}) and (\ref{eq:BestSS}) and observe the numerical results in Fig.~\ref{fig:BetaEst}, we can see that the ``predicted" number of infectious people is about half the ``real" number of infectious people. 

To determine if we see underestimation in a more general scenario, we compare (\ref{eq:BestSS}) analytically to our delay steady state, (\ref{eq:Idss}), also with $\alpha \neq 0$. For simplicity, we Taylor expand to find the leading order term in $\mu$, assuming that $\mu \ll  \alpha, \gamma, \sigma$ since $\frac{1}{\mu}$ represents the length of a lifetime in days and is much longer than the length of quarantine, infectious, and exposed periods.
 
This leads to $I^*_{D} \approx  \frac{\mu}{\alpha}\left(1-e^{-\alpha/\gamma}\right)-\frac{\mu}{\beta_{\textnormal{real}}}$ and $ I^*_{\textnormal{est}} \approx  \frac{\mu}{\gamma}\left(1-\frac{\alpha}{\alpha + \gamma}\right) - \frac{\mu}{\beta_{\textnormal{real}}} $. Thus comparing the ``real'' and ``predicted'' infection levels requires comparing factor $C_D=1-e^{-\alpha/\gamma}$ to factor $C_\textnormal{est}=\frac{\alpha}{\gamma}\left(1-\frac{\alpha}{(\alpha+\gamma)} \right)$. 

At this point, it is still difficult to tell which is greater, but we can conclude that the extent of the discrepancy between the ``real" system and the ``predicted" system depends on the quarantine rate and the recovery rate. 

If we further assume a small quarantine rate such that $\alpha \ll  \gamma$, we can expand the two steady states further and find that $C_{D} \approx \frac{\alpha}{\gamma}\left(1-\frac{\alpha}{2\gamma}\right)$ and $ C_{\textnormal{est}} \approx \frac{\alpha}{\gamma}\left(1-\frac{\alpha}{\gamma}\right)$. In this limit, $C_D>C_\textnormal{est}$.

Therefore, at least for low quarantine, it can be seen that the ``predicted" exponential system's infectious steady state is less than the ``real" delay system's infectious steady state, confirming that the exponential system will underestimate the number of infectious in our model when a control measure is applied.


\section{Effect of Number of Compartments on Infectious Steady State}
\label{CompartSS}

We can use the multi-infected compartment model to interpolate between the exponential system and the delay system by varying the number of compartments. Now we want to investigate how this variation will affect the infectious steady state. It is important to know if a particular method of modeling tends to overestimate or underestimate the infectious steady state when attempting to apply accurate control measures. We begin by taking (\ref{eq:Iss}) and letting the number of exposed compartments, $n$, increase to infinity while maintaining only one infectious compartment and one quarantine compartment, $m = 1$. This allows the distribution of the exposed period to move from an exponential distribution to a constant period while the infectious period remains exponentially distributed. We will be able to see if the narrower distribution of the exposed class increases or decreases the infectious steady state, impacting the needed amount of control. We then repeat this analysis for the infectious class, allowing the number of infectious compartments, $m$, to increase to infinity while maintaining only one exposed compartment. 

\begin{figure}[t]
\begin{center}
\includegraphics[scale=.35]{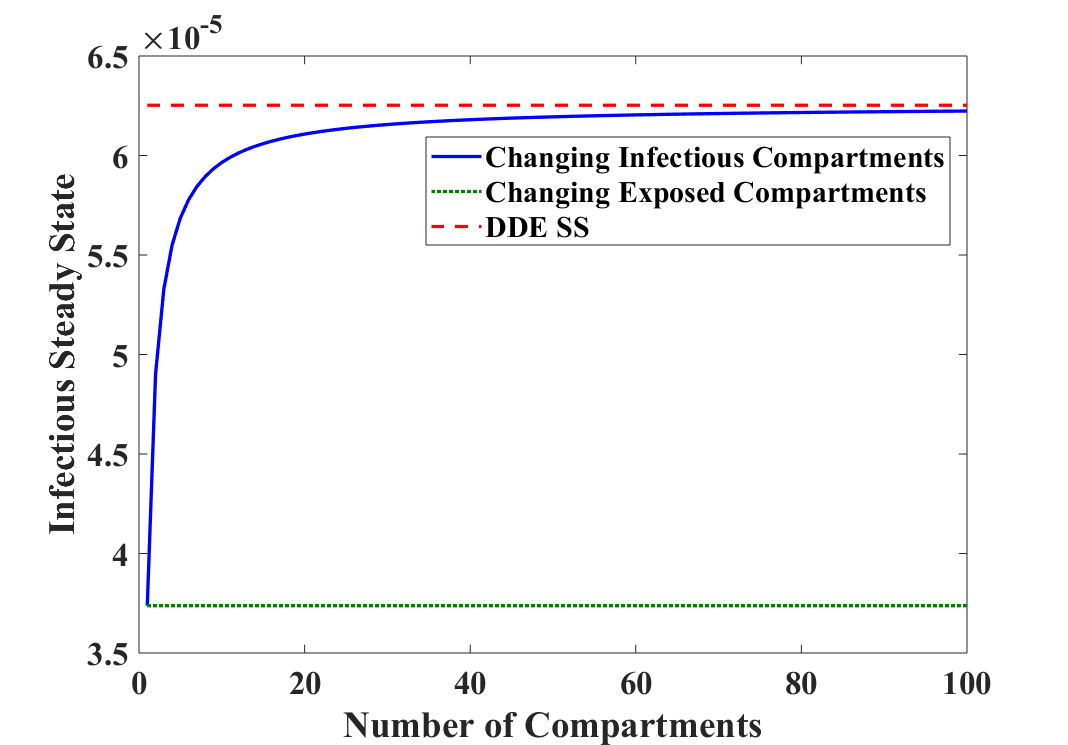} 
\end{center}
\caption{Infectious steady state of the multi-infected model as the number of exposed (dotted curve) or infectious (solid curve) compartments increases. The infectious steady state of the delay system is shown for comparison (dashed curve)  \label{fig:mterm}}
\end{figure}

In Fig.~\ref{fig:mterm}, using parameters from Table \ref{table:1}, we can see that as the number of infectious compartments increases, the multi-infected system's infectious steady state approaches the delay system's steady state. However, as the number of exposed compartments increases, the multi-infected system's infectious steady state seems to stay constant.

We now analytically investigate this result in order to form a more general argument beginning with the change in the number of exposed compartments. Since we are only changing $n$, we focus on $C_E$ (\ref{eq:CE}), which when increased, increases the infectious steady state. 

When $n = 1$,  the exposed class has an exponential distribution and $C_E$ becomes
\begin{equation}
\label{eq:enterm}
\frac{1}{\frac{\mu}{\sigma}+1}.
\end{equation}
As $n$ approaches infinity, the distribution of time spent exposed approaches a delta distribution and $C_E$ 
becomes
\begin{equation}
\label{eq:dnterm}
e^{-\mu/\sigma}.
\end{equation}

Next, we assume that $\frac{\mu}{\sigma} \ll 1$ due to the much smaller birth rate. We use a Taylor expansion to the second order on (\ref{eq:enterm}) about $\frac{\mu}{\sigma}$, resulting in
\begin{equation}
\label{eq:teenterm}
1-\frac{\mu}{\sigma} + \left(\frac{\mu}{\sigma}\right)^2.
\end{equation}
The Taylor expansion for (\ref{eq:dnterm}) about $\frac{\mu}{\sigma}$ results in
\begin{equation}
\label{eq:tednterm}
1-\frac{\mu}{\sigma} + \left(\frac{1}{2}\right)\left(\frac{\mu}{\sigma}\right)^2.
\end{equation}
Comparing the two expanded terms, we see that $C_E$ using an exponential distribution is greater than $C_E$ using a delay distribution, or 
\[
\frac{1}{\frac{\mu}{\sigma}+1} > e^{-\mu/\sigma}.
\]
Therefore, using DDEs (or a delta distribution) for the exposed class will decrease the infectious steady state. However, comparing this to our numerical results in Fig.~\ref{fig:mterm}, we observe a negligible decrease in the infectious steady state.

Next, we perform the same analysis for varying the number of infectious compartments, $m$. This time, we focus on $C_I$ (\ref{eq:CI}), which when increased, also increases the infectious steady state.

Again, we can assume $\frac{\mu}{\gamma} \ll 1$ due to difference in time scales. For convenience, we make an extra, potentially unrealistic, assumption that the quarantine rate $\alpha$ is much smaller than the recovery rate $\gamma$. Letting $m=1$ gives us an exponential distribution in the infectious compartment and $C_I$ 
becomes
\begin{equation}
\label{eq:emterm}
1-\left(\frac{1}{1+\frac{\mu+\alpha}{\gamma}}\right)
\end{equation}
with a Taylor expansion to the second order about small $\frac{\mu+\alpha}{\gamma}$ of 
\begin{equation}
\label{eq:teemterm}
\frac{\mu+\alpha}{\gamma} - \left(\frac{\mu+\alpha}{\gamma}\right)^2.
\end{equation}
As $m$ approaches infinity, the infectious class approaches a delta distribution and $C_I$ 
becomes
\begin{equation}
\label{eq:dmterm}
1-e^{-(\mu+\alpha)/\gamma}
\end{equation}
with a Taylor expansion to the second order about small $\frac{\mu+\alpha}{\gamma}$ of
\begin{equation}
\label{eq:tedmterm}
\frac{\mu+\alpha}{\gamma} - \left(\frac{1}{2}\right)\left(\frac{\mu+\alpha}{\gamma}\right)^2.
\end{equation}
Comparing the two expanded terms, we see that $C_I$ using a delta distribution is greater than $C_I$ using an exponential distribution, or 
\[
1-\frac{\gamma}{\alpha+\gamma +\mu} < 1- e^{-(\mu+\alpha)/\gamma}. 
\]
Thus, using DDEs for the infectious class will increase the infectious steady state.

Numerically, we have shown that increasing the number of infectious compartments allows the infectious steady state to reach more accurate levels, while increasing the number of exposed compartments has little impact. When investigating this analytically, using a time delay in the exposed compartment appears to decrease the infectious steady state. For small control levels, we showed analytically that using a time delay in the infectious compartment increases the infectious steady state. 

Together, these results reinforce that using an exponential distribution can lead to underestimates in predicted steady states, with new emphasis on using the correct distribution in the infectious class.

\section{Effect of Length of Time Delay on Infectious Steady State}
\label{DelayLength}

We now investigate whether the length of the time delay has an impact on the discrepancy between the exponential and the delay systems' infectious steady states. We test this by focusing on each time delay separately.

We first look at the time delay in the exposed class, varying $\tau_1$ while assuming the infectious period and number of infectious compartments are fixed. Therefore, we again only need to focus on $C_E$ (\ref{eq:CE}). To study the difference between the steady state of the exponential and delay system, we need to use the expanded form of (\ref{eq:enterm}) and (\ref{eq:dnterm}) for slow birth rate. Subtracting (\ref{eq:tednterm}) from (\ref{eq:teenterm}) results in
\[
\left(\frac{1}{2}\right)\left(\frac{\mu}{\sigma}\right)^2 .
\]
Thus the exponential system's infectious steady exceeds that of the delay system by
\begin{equation}
\label{eq:varyt1}
\frac{\mu}{\alpha+\mu}C_I \frac{1}{2}\left(\mu\tau_1\right)^2
\end{equation}
 since $\tau_1 = \frac{1}{\sigma}$.  (Recall that $C_I$ is independent of $\sigma$ and $\tau_1$.)
As the time delay in the exposed class increases, the discrepancy between the infectious steady state of the exponential and the delay system increases.

\begin{figure}[t]
\subfloat[]{\label{fig:t1dis}\includegraphics[scale=.3]{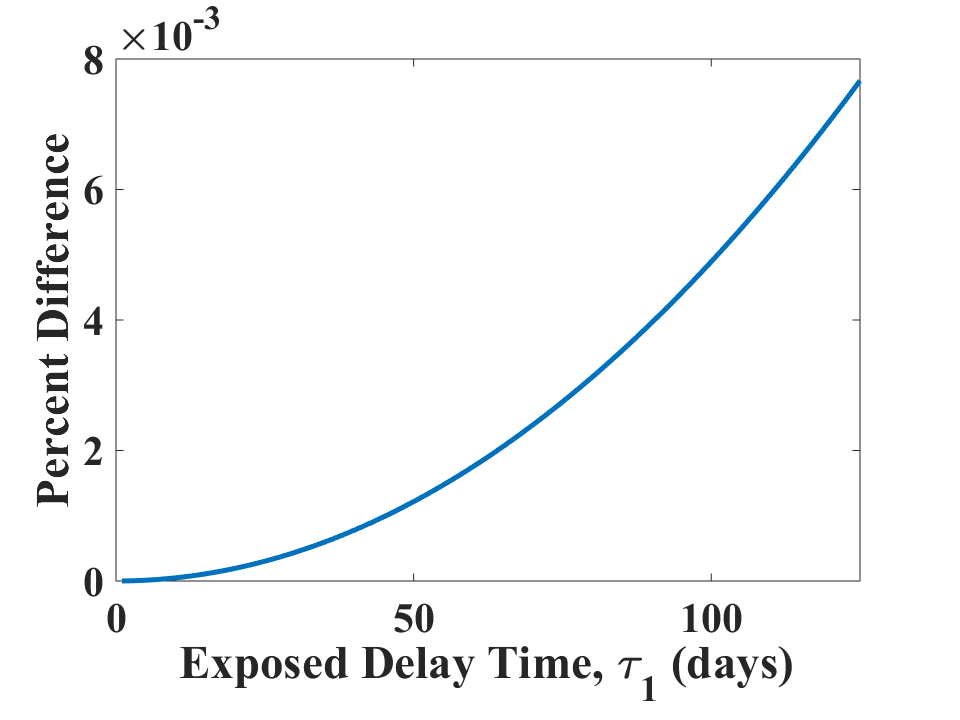}}
\subfloat[]{\label{fig:t2dis}\includegraphics[scale=.30]{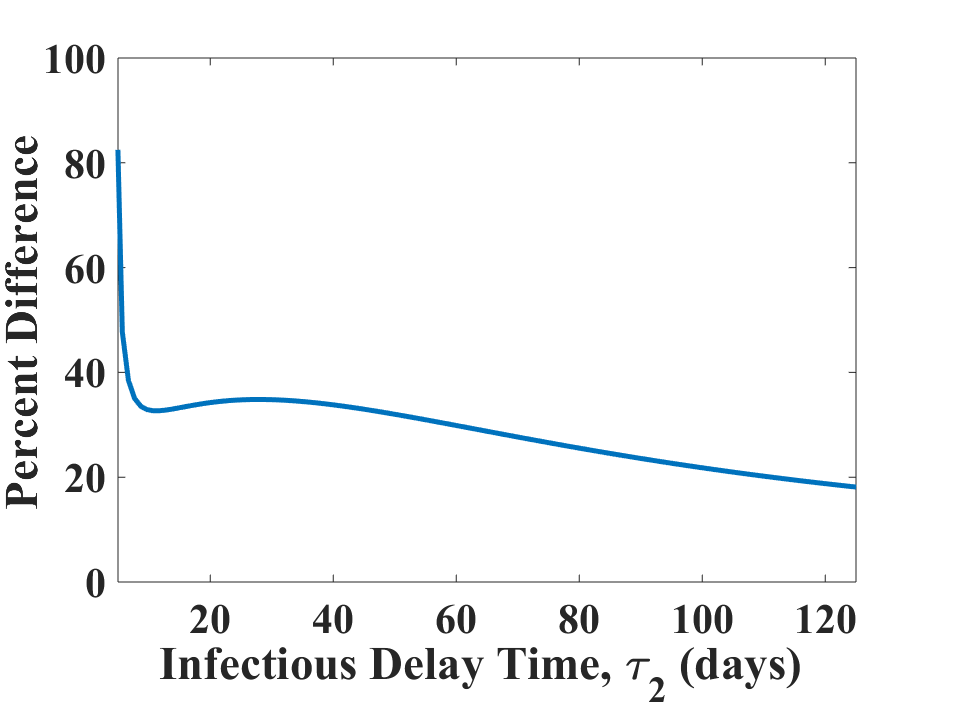}}
\caption{(a) Percent difference in infectious steady state between using an exponential distribution and delay distribution in the exposed class versus time spent exposed, $\tau_1$. (b) Percent difference between using an exponential distribution and delay distribution in the infectious class versus time spent infectious, $\tau_2$}
\label{fig:Discrepancies}
\end{figure}

In Fig.~\ref{fig:t1dis} we plot the percent difference in the infectious steady state between using an exponential distribution and delay distribution in the exposed class as $\tau_1$ increases. We maintain a time delay in the infectious compartment for consistency. We see that the percent difference is quite small, implying that the time delay in the exposed class has little impact on the difference in infectious steady states. 

Note that in Fig.~\ref{fig:t1dis} we stop $\tau_1$ at 123 days because of the existence of a Hopf bifurcation in the delay system at approximately $\tau_1 = 123.75$, after which the infectious steady state is less relevant. In order to locate the Hopf bifurcation, we tested different values for $\tau_1$ and used dde23 in Matlab to numerically integrate (\ref{eq: dde}) from an initial condition that included a slight perturbation from the endemic steady state. For $\tau_1 < 123.5$, the system approached the endemic steady state. For $\tau_1 > 123.75$, the system continued to oscillate with increasing amplitude.

 A similar approach to finding the discrepancy between the two steady states by varying the delay $\tau_2$ in the infectious compartment would require us to focus on $C_I$ (\ref{eq:CI}).  However, we cannot make simplifying assumptions in a convenient limit because if $\tau_2$ is long, the assumption that the quarantine rate, $\alpha$, is slow compared to the recovery rate $\gamma$ (made in the expansion of $C_I$ in (\ref{eq:teemterm}) and (\ref{eq:tedmterm})) does not hold, while if the infectious period $\tau_2$ is too short, the system will approach the disease free state.

Therefore, we resort to numerics.  The percent difference between using an exponential distribution and delay distribution for the infectious class is plotted in Fig \ref{fig:t2dis}. Again, a time delay was used in the exposed class for consistency. This time, we see that the delay in the infectious class has a much greater impact on the difference between the infectious steady states and that smaller time delays contribute to greater differences.

Note that by numerically searching for a Hopf bifurcation, we determined that there was no Hopf bifurcation for $\tau_2 < 125$. However, the system appears to be below epidemic threshold at $\tau_2 = 4.25$ days, so we only plot for $\tau_2\geq 5$ days.

\section{Effects of Different Control Applications}
\label{DiffControls}

We next consider other control strategies that are applied to different subsets of the population and whether that has an impact on if time delays must be used. For instance, would applying a vaccine to the susceptible population lead to different results than applying quarantine to an exposed population when time delays were used?

\begin{figure}[t]
\subfloat[]{\label{fig:ChangeNoQ}\includegraphics[width = 0.5 \textwidth]{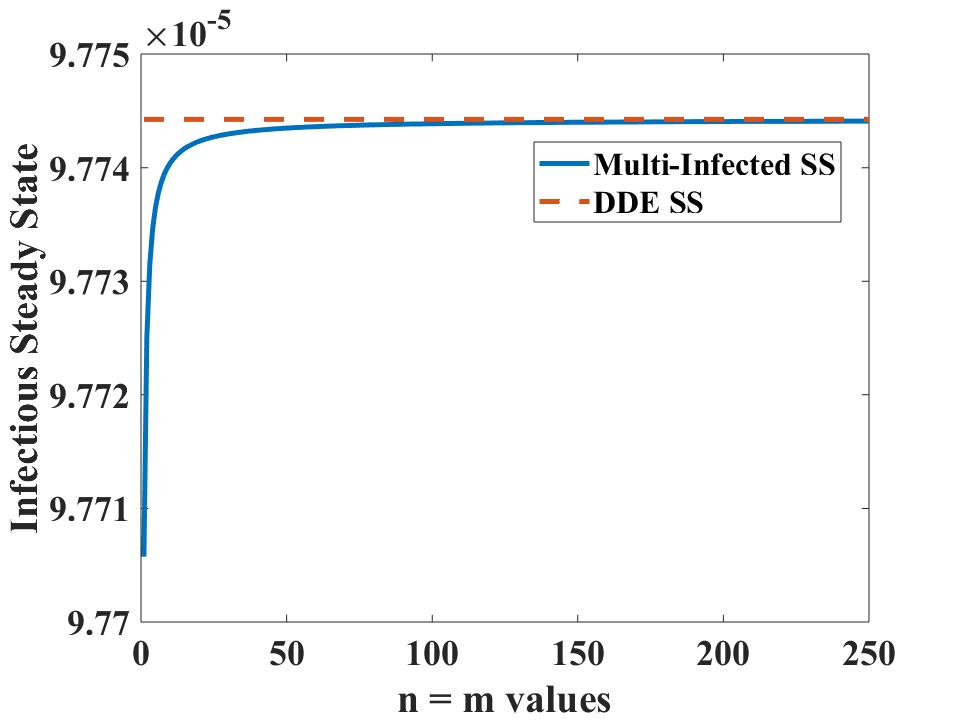}}
\subfloat[]{\label{fig:QonE}\includegraphics[width = 0.5 \textwidth]{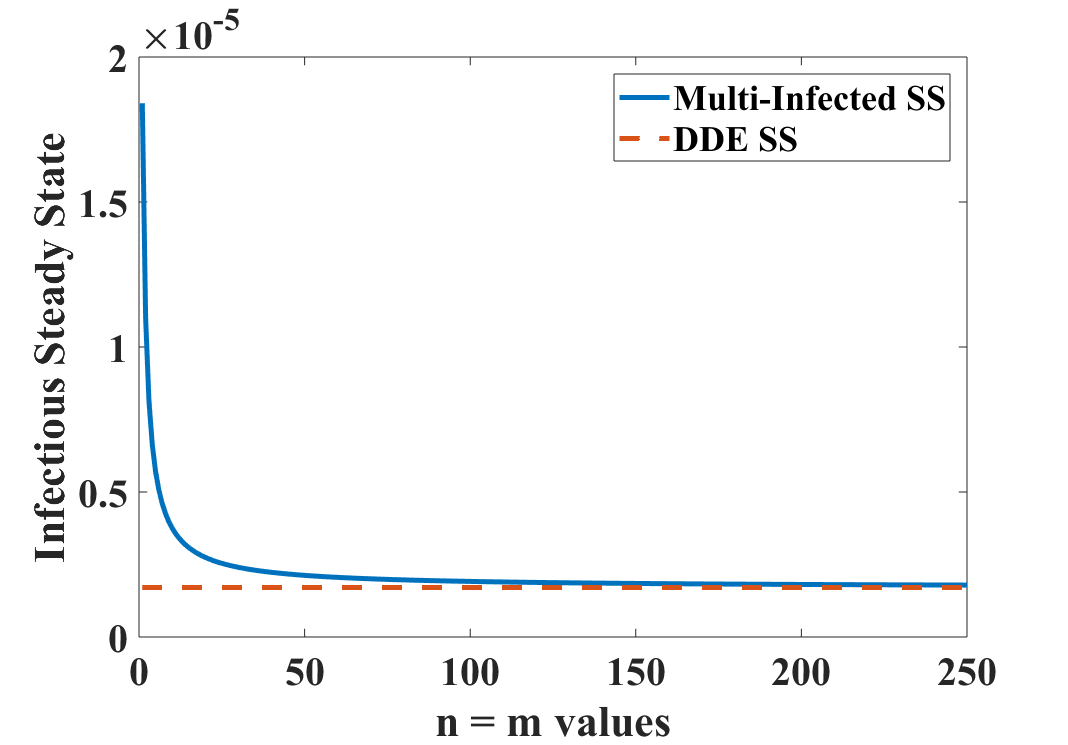}}
\caption{(a) Infectious steady state of a system without any control measure as the number of infectious compartments, $m$, and the number of exposed compartments, $n$ increases at the same rate. (b) Infectious steady state when quarantine is applied to the exposed class as the number of infectious compartments, $m$, and the number of exposed compartments, $n$ increases at the same rate. The same quarantine rate is used on the exposed class as was used on the infectious class. See Table \ref{table:1}}
\label{fig:OtherControls}
\end{figure}

We first investigate the case where there was no control. For our analysis, we use our models with $\alpha = 0$ and parameters from Table \ref{table:1}. As we already saw with infectious quarantine control, the exponential model underestimates the infectious steady state as compared with the delay model.  As we increase the number of exposed and infectious compartments, the multi-infected infectious steady state (see (\ref{eq:Iss}) for $\alpha = 0$) increases to approach the delay system's infectious steady state (see Fig.~\ref{fig:ChangeNoQ}). Similar to the case where quarantine was applied to the infectious class, the number of exposed compartments has little impact on changing the multi-infected system's infectious steady state, while increasing the number of infectious compartments helps increase the multi-infected infectious steady state to match the delay system (data not shown).

We observe a similar scenario when applying a vaccine to the susceptible class. In this model, a proportion of the susceptible population is transferred at vaccination rate $\nu$ to the recovered/removed class (see Fig.~\ref{fig:compflowV}). Steady states are given in Appendix \ref{VonS}.  Again, the exponential model underestimates the infectious steady state (data not shown).  As we increase the number of infectious compartments, the multi-infected infectious steady state also increases to approach the delay steady state. Increasing the number of exposed compartments has little effect on the infectious steady state.
 
 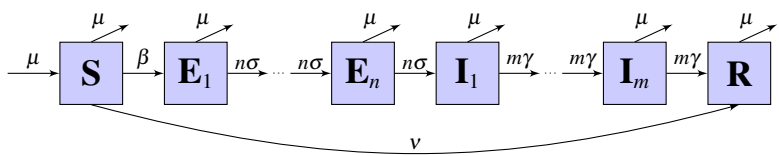
\begin{figure}[t]
\begin{center}
   \begin{tikzpicture}[node distance=2.5cm,auto,>=latex',scale=.55, every node/.style={scale=.55}]
    \node [int] (S) {\Huge \textbf{S}};
    \node (muin) [left of=S,node distance = 2. cm, coordinate] {$\mu$};
    \node [int] (E1) [right of=S] { \Huge $\textnormal{\textbf{E}}_{\scriptscriptstyle 1}$};
    \node [fill=white, minimum size=.5cm] (dots1) [right of =E1,node distance = 2 cm] {$\cdots$};
    \node [int] (En) [right of=dots1,node distance = 2 cm] { \Huge $\textnormal{\textbf{E}}_{\scriptscriptstyle n}$};
    \node [int] (I1) [right of=En] { \Huge $\textnormal{\textbf{I}}_{\scriptscriptstyle 1}$};
    \node [fill=white, minimum size=.5cm] (dots2) [right of =I1,node distance = 2 cm] {$\cdots$};
    \node [int] (Im) [right of=dots2,node distance = 2 cm] { \Huge $\textnormal{\textbf{I}}_{\scriptscriptstyle m}$};
    \node [int] (R) [right of=Im] {\Huge \textbf{R}};

    \path[->] (muin) edge node {\Large $\mu$} (S);
    \path[->] (S) edge node {\Large $\beta$} (E1);
    \path[->] (E1) edge node {\Large $n\sigma$} (dots1);
    \path[->] (dots1) edge node {\Large $n\sigma$} (En);
    \draw[->] (En) edge node {\Large $n\sigma$} (I1) ;
    \draw[->] (I1) edge node {\Large $m\gamma$} (dots2) ;
    \draw[->] (dots2) edge node {\Large $m\gamma$} (Im) ;
    \draw[->] (Im) edge node {\Large $m\gamma$} (R) ;

    \draw[->] (S.90) --++ (25:1cm) node[midway]{\Large $\mu$};
    \draw[->] (E1.90) --++ (25:1cm) node[midway]{\Large $\mu$};
    \draw[->] (En.90) --++ (25:1 cm) node[midway]{\Large $\mu$};
	\draw[->] (I1.90) --++ (25:1 cm) node[midway]{\Large $\mu$};
    \draw[->] (Im.90) --++ (25:1 cm) node[midway]{\Large $\mu$};
    \draw[->] (R.90) --++ (25:1 cm) node[midway]{\Large $\mu$};
     \path[->] (S.south) edge[bend right = 15] node  {\Large $\nu$} (R.south);
    
\end{tikzpicture} 
\end{center}
\caption{Flow chart for multi-infected compartment model with a vaccine applied to the susceptible class, where $n$ is the number of exposed compartments, $m$ is the number of infectious compartments, and $\nu$ is the vaccination rate.  Other parameters are as before. See Table \ref{table:1} \label{fig:compflowV}}
\end{figure}

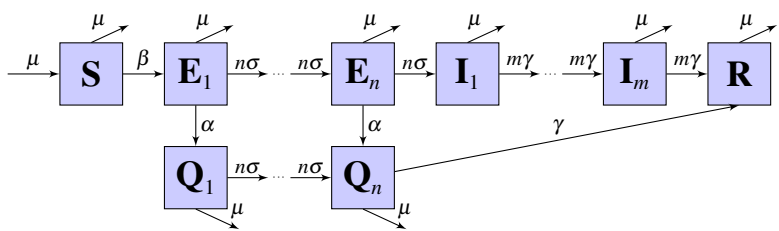
\begin{figure}[t]
\begin{center}

 \begin{tikzpicture}[node distance=2.5cm,auto,>=latex',scale=.55, every node/.style={scale=.55}]
    \node [int] (S) {\Huge \textbf{S}};
    \node (muin) [left of=S,node distance = 2. cm, coordinate] {$\mu$};
    \node [int] (E1) [right of=S] { \Huge $\textnormal{\textbf{E}}_{\scriptscriptstyle 1}$};
    \node [fill=white, minimum size=.5cm] (dots1) [right of =E1,node distance = 2 cm] {$\cdots$};
    \node [int] (En) [right of=dots1,node distance = 2 cm] { \Huge $\textnormal{\textbf{E}}_{\scriptscriptstyle n}$};
    \node [int] (I1) [right of=En] { \Huge $\textnormal{\textbf{I}}_{\scriptscriptstyle 1}$};
    \node [fill=white, minimum size=.5cm] (dots2) [right of =I1,node distance = 2 cm] {$\cdots$};
    \node [int] (Im) [right of=dots2,node distance = 2 cm] { \Huge $\textnormal{\textbf{I}}_{\scriptscriptstyle m}$};
    \node [int] (R) [right of=Im] {\Huge \textbf{R}};
    \node [int] (Q1) [below of=E1, node distance = 2.5 cm] { \Huge $\textnormal{\textbf{Q}}_{\scriptscriptstyle 1}$};
    \node [fill=white, minimum size=.5cm] (dots3) [right of =Q1,node distance = 2 cm] {$\cdots$};
    \node [int] (Qn) [below of=En, node distance = 2.5 cm] { \Huge $\textnormal{\textbf{Q}}_{\scriptscriptstyle n}$};

    \path[->] (muin) edge node {\Large $\mu$} (S);
    \path[->] (S) edge node {\Large $\beta$} (E1);
    \path[->] (E1) edge node {\Large $n\sigma$} (dots1);
    \path[->] (dots1) edge node {\Large $n\sigma$} (En);
    \draw[->] (En) edge node {\Large $n\sigma$} (I1) ;
    \draw[->] (I1) edge node {\Large $m\gamma$} (dots2) ;
    \draw[->] (dots2) edge node {\Large $m\gamma$} (Im) ;
    \draw[->] (Im) edge node {\Large $m\gamma$} (R) ;
    \draw[->] (E1) edge node {\Large $\alpha$} (Q1);
    \draw[->] (Q1) edge node {\Large $n\sigma$} (dots3) ;
    \draw[->] (dots3) edge node {\Large $n\sigma$} (Qn) ;
    \draw[->] (En) edge node {\Large $\alpha$} (Qn) ;
    \draw[->] (Qn) edge node[midway] {\Large $\gamma$} (R.-90) ;

    \draw[->] (S.90) --++ (25:1cm) node[midway]{\Large $\mu$};
    \draw[->] (E1.90) --++ (25:1cm) node[midway]{\Large $\mu$};
    \draw[->] (En.90) --++ (25:1 cm) node[midway]{\Large $\mu$};
	\draw[->] (I1.90) --++ (25:1 cm) node[midway]{\Large $\mu$};
    \draw[->] (Im.90) --++ (25:1 cm) node[midway]{\Large $\mu$};
    \draw[->] (R.90) --++ (25:1 cm) node[midway]{\Large $\mu$};
    \draw[->] (Q1.-90) --++ (-25:1.15 cm) node[pos = 0.7]{\Large $\mu$};
    \draw[->] (Qn.-90) --++ (-25:1.15 cm) node[pos = 0.7]{\Large $\mu$};
    
\end{tikzpicture} 
\end{center}
\caption{Flow chart for multi-infected compartment model with a quarantine applied to the exposed class, where $n$ is the number of exposed/quarantine compartments and $m$ is the number of infectious compartments. Other parameters are as before. See Table \ref{table:1} \label{fig:compflowQonE}}
\end{figure}

Lastly, we investigate what happens when we apply quarantine to the exposed class rather than to the infectious class. Individuals are still quarantined at rate $\alpha$ and progress through the $n$ quarantine compartments (see Fig.~\ref{fig:compflowQonE}). Assuming they are being cured while in the quarantine class, they move straight into the recovered class after $n$ quarantine compartments. The infectious steady state is given in Appendix \ref{SSQonE}.  In contrast to the other control strategies we considered, quarantine of exposed individuals can lead to the exponential system overestimating the infectious steady state as compared with the delay system. For the parameters studied, the multi-infected compartment's infectious steady state actually decreases to approach a smaller delay system infectious steady state (see Fig.~\ref{fig:QonE}). Also, the number of exposed compartments contributes mostly to the decrease in infectious steady state where the number of infectious compartments has little impact on the change in infectious steady state. This implies that a time delay would be more beneficial in the exposed class rather than the infectious class as seen when applying a quarantine to the infectious class. 

In a similar observation made by \citet{wearing2005}, adding more compartments to the exposed class slows down the movement of exposed individuals who would otherwise quickly move into the infectious class.  This allows for more of them to become quarantined and results in fewer of them spreading the disease.When quarantine was applied to the infectious class, there were still individuals spreading the disease before having a chance to become quarantined. Adding more compartments slows down the movement of infectious individuals into the recovered class, allowing for more disease spread and, therefore, increasing the infectious steady state.  

\section{Conclusion}
\label{Conclusion}

Mathematical models play an important role in predicting the size of an infectious population during an epidemic outbreak. The size of the infectious population is key for allocating resources, time, and money. As explained by  \citet{meltzer2016}, some mathematical models need to be produced and evaluated under a strict time constraint.  Therefore it can be preferable to use the quickest and easiest method even if it sacrifices some accuracy. The exponential system is favorable due to its ease, but if mathematicians are not careful, the parameters used could lead to a significant underestimation of the infectious steady state as highlighted in \citet{wearing2005}. We also observed an underestimation of the infectious steady state of our exponential model when fitting the transmission rate and applying quarantine of infectious individuals. 
This underestimation could lead to inaccuracies in the amount of control recommended, which could lead to difficulty in eradicating the epidemic. 

To explore in more detail the role of the distribution used for the time spent exposed or infectious, we created a multi-infected compartment model to interpolate between our exponential model and our delay model and explored whether a particular time delay had a greater impact on the infectious steady state, which could help to make an epidemic model more simple and accurate. We found that using a time delay in the exposed class led to a smaller infectious steady, but with the parameters used in \citet{hu2015}, this decrease was negligible. When using a time delay in the infectious class, the infectious steady state increased and, with our parameter values, the increase was significant. This allowed us to conclude that having a time delay in the infectious class has more of an impact on the infectious steady state than having a time delay in the exposed class.

Next, we tested how the magnitude of the time delay affected the infectious steady state. Some diseases will have shorter exposed and infectious periods such as measles and influenza \citep{hethcote2000,wearing2005,safi2011}, and others will have very large periods such as HIV \citep{anderson1992}. It is important to consider the length of a time delay when selecting the type of distribution needed for a mathematical model. We found that as time spent infectious increases (a larger $\tau_2$), the difference between the two systems' infectious steady states decreases. Therefore, if the disease has a large infectious period, the infectious steady state would be similar for the exponential system and the delay system. In this case, it would be best to use the exponential system due to its simplicity. When doing a similar analysis for the magnitude of the exposed time delay, we saw that the difference between our two infectious steady states was extremely small. This again emphasizes the importance of the time delay in the infectious class versus the exposed class. 

Finally, we investigated how the placement of a control measure affected the impact each time delay had on the infectious steady state. A disease may have multiple possible control measures available. For instance, Ebola does not yet have a vaccine, but it is important to consider quarantining the exposed population, promoting safe burial practices, and hospitalization rates \citep{hu2015, haas2014, pandey2014, meltzer2014estimating, meltzer2016}. For this study, we looked at how vaccinating the susceptible class or quarantining the exposed or infectious class would affect the model. We found that placing a vaccination on the susceptible class led to similar results as having a quarantine on the infectious class. The infectious time delay, $\tau_2$, had a greater impact on the infectious steady state. On the contrary, when a quarantine was applied to the exposed class, the exposed time delay, $\tau_1$, led to a decrease in the infectious steady state as it approached the delay's infectious steady state. Although the distribution of infectious periods is usually the most important because infectives spread the disease, we suggest that if a control strategy acts directly on a compartment, the distribution of time spent in that compartment becomes similarly important.  To understand the impact of a control strategy, it may be necessary to model the timing of that control in detail rather than assuming it is exponentially distributed.

We believe caution must be taken when applying control measures to models using ordinary differential equations. Although simple and quick to analyze, they could lead to significant underestimates or even overestimates in the prediction of the infectious population, causing difficulties in controlling the disease. Future modeling may include yet more types of control measures, such as hospitalization rates and burial practices, along with incorporating death by disease, to better elucidate the role of time distributions spent in each compartment.

\section{Appendix}
\appendix

\section{Endemic Steady State for Exponential System}
\label{ExpSS}
The endemic steady state for (\ref{eq: exp}) is
\begin{align*}
S_E^* &= \frac{(\sigma + \mu)(\alpha +\gamma + \mu)}{\beta\sigma} \\
E_E^* &= \frac{-\mu(\sigma + \mu)(\alpha +\gamma + \mu)}{\beta\sigma(\sigma+\mu)} + \frac{\mu}{\sigma+\mu} \\
I_E^* &=\frac{\mu}{\alpha+\mu} \left(\frac{\sigma}{\sigma+\mu}\right)\left[1-\left(\frac{\gamma}{\gamma +\alpha+ \mu}\right)\right] -\frac{\mu}{\beta}\\
Q_E^* &= \frac{-\alpha\mu}{\beta(\gamma+\mu)} + \frac{\sigma\alpha\mu}{(\gamma+\mu)(\sigma+\mu)(\alpha+\gamma+\mu)} \\
R_E^* &= \frac{-\gamma(\alpha+\gamma+\mu)}{\beta(\gamma+\mu)} + \frac{\gamma\sigma}{( \gamma+\mu )(\sigma+\mu)}
\end{align*}
where our main focus is on $I_E^*$.

\section{Endemic Steady State for Delay System}

\label{DDESS}
The endemic steady state for (\ref{eq: dde}) is
\begin{align*}
S_D^* &= \frac{e^{\mu\tau_1+(\alpha+\mu)\tau_2}(\alpha + \mu)}{\beta \left(e^{(\alpha+\mu)\tau_2}-1\right)} \\
E_D^* &= \frac{\left(e^{-\mu\tau_1}-1\right)\left((\alpha+\mu)e^{\mu\tau_1 + (\alpha+\mu)\tau_2}-\beta\left(e^{(\alpha+\mu)\tau_2}-1\right)\right)}{\beta\left(e^{(\alpha+\mu)\tau_2}-1\right)}\\
I_D^* &= \frac{\mu}{\alpha+\mu} e^{-\mu\tau_1}\left[1-e^{-(\mu+\alpha)\tau_2}\right] -\frac{\mu}{\beta} \\
Q_D^* &= \frac{e^{-\mu\tau_1-(\alpha+\mu)\tau_2}\left[(\alpha+\mu)e^{\mu\tau_1+(\alpha+\mu)\tau_2}-\beta\left(e^{(\alpha+\mu)\tau_2}-1\right)\right]\left[\alpha e^{\alpha\tau_2}\left(1-e^{\mu\tau_2}\right)+\mu\left(e^{\alpha\tau_2}-1
\right)\right]}{\beta(\alpha+\mu)\left(e^{(\alpha+\mu)\tau_2}-1\right)} \\
R_D^* &= \frac{\beta\left(1-e^{\tau_2(\alpha+\mu)}\right)+(\alpha+\mu)e^{\mu\tau_1+\tau_2(\alpha+\mu)}}{\beta e^{\mu(\tau_1+\tau_2)}[1-e^{\tau_2(\alpha+\mu)}]}
\end{align*}
where our main focus is on $I_D^*$. Recall $\tau_1 = \frac{1}{\sigma}$ and $\tau_2 = \frac{1}{\gamma}$. 

\section{$R_0$ for Multi-Infected Compartment Model}
\label{R0MIC}
Using the method established by \cite{van2002}, we find the basic reproduction number, $R_0$, for the multi-infected compartment model, (\ref{eq: comp}). First, note that there are $n+2m$ infected compartments: $E_1, E_2, ..., E_n, I_1, I_2, ..., I_m,$ and $Q_1, Q_2, ..., Q_m$. Therefore, $(n+2m) \times (n+2m)$ matrices are needed such that $\mathscr{F}$ is the new infections matrix and $\mathscr{V}$ holds the negative of all other terms. Using all equations of the system, we obtain

\[ \mathscr{F} = 
\begin{pmatrix}
0 \\
BS\sI \\
0 \\
\vdots \\
0 \\
0 \\
0 \\
\vdots \\
0 \\ 
0 \\
0 \\
\vdots \\
0 \\
0 \\
 \end{pmatrix}  
\]

and 

\[ \mathscr{V} = 
\begin{pmatrix}
  -\mu + BS\sI + \mu S \\
  (\mu + n\sigma) E_1\\
  -n\sigma E_1  + (\mu +n\sigma)E_2 \\
  \vdots \\
  -n\sigma E_{n-1} + (\mu +n\sigma) E_n \\
  -n\sigma E_n + (\mu +\alpha+m\gamma) I_1 \\
  -m\gamma I_1 + (\mu+\alpha + m\gamma) I_2 \\
  \vdots\\
  -m\gamma I_{m-1} + (\mu + \alpha +m\gamma) I_m \\
  -\alpha I_1 +(\mu+m\gamma) Q_1 \\
  -\alpha I_2 -m\gamma Q_1 + (\mu +m\gamma)Q_2 \\
  \vdots \\
  -\alpha I_m -m\gamma Q_{m-1} +(\mu+m\gamma)Q_m \\
  -\gamma(I_m+Q_m) +\mu R
 \end{pmatrix}.  
\]

Next, using only the infected compartments, the derivative submatrices at the disease free equilibrium are
calculated to be

\[ F = 
\begin{pmatrix}
0 & 0 & \cdots & 0 & \beta & \beta & \cdots & \beta & 0 & 0 & \cdots & 0 \\
0 & 0 & \cdots & 0 & 0 & 0 & \cdots & 0 & 0 & 0 & \cdots & 0 \\
\vdots & \vdots & \ddots & \vdots & \vdots & \vdots & \ddots & \vdots & \vdots & \vdots & \ddots & \vdots \\
0 & 0 & \cdots & 0 & 0 & 0 & \cdots & 0 & 0 & 0 & \cdots & 0  
 \end{pmatrix}  
\]

and

{\tiny
\[ 
V = 
\begin{pmatrix}
\mu +n\sigma & 0 & 0 & \cdots & 0 & 0 & 0 & \cdots & 0 & 0 & 0 & \cdots & 0 \\
-n\sigma & \mu+n\sigma & 0 & \cdots & 0 & 0 & 0 & \cdots & 0 & 0 & 0 & \cdots & 0 \\
0 & -n\sigma & \mu+n\sigma & \cdots & 0 & 0 & 0 & \cdots & 0 & 0 & 0 & \cdots & 0 \\
\vdots & \vdots & \vdots & \ddots & \vdots & \vdots & \vdots & \ddots & \vdots & \vdots & \vdots & \ddots & \vdots \\
0 & 0 & 0 & \cdots & \mu+n\sigma & 0 & 0 & \cdots & 0 & 0 & 0 & \cdots & 0 \\
0 & 0 & 0 & \cdots & -n\sigma & \mu+\alpha+m\gamma & 0 & \cdots & 0 & 0 & 0 & \cdots & 0 \\
0 & 0 & 0 & \cdots & 0 & -m\gamma & \mu+\alpha+m\gamma & \cdots & 0 & 0 & 0 & \cdots & 0 \\
\vdots & \vdots & \vdots & \ddots & \vdots & \vdots & \vdots & \ddots & \vdots & \vdots & \vdots & \ddots & \vdots \\
0 & 0 & 0 & \cdots & 0 & 0 & 0 & \cdots & \mu+\alpha+m\gamma & 0 & 0 & \cdots & 0 \\
0 & 0 & 0 & \cdots & 0 & -\alpha & 0 & \cdots & 0 & \mu+m\gamma & 0 & \cdots & 0 \\
0 & 0 & 0 & \cdots & 0 & 0 & -\alpha & \cdots & 0 & -m\gamma & \mu+m\gamma & \cdots & 0 \\
\vdots & \vdots & \vdots & \ddots & \vdots & \vdots & \vdots & \ddots & \vdots & \vdots & \vdots & \ddots & \vdots \\
0 & 0 & 0 & \cdots & 0 & 0 & 0 & \cdots & -\alpha & 0 & 0 & \cdots & \mu+m\gamma \\
 \end{pmatrix}.  
\]
}

Then taking the product of $F$ and $V^{-1}$ yields an upper triangular matrix of
{\tiny
\[ FV^{-1} = 
\begin{pmatrix}
\beta \left(\frac{n\sigma}{\mu+n\sigma}\right)^n\left(\frac{1}{\mu+\alpha+m\gamma}\right)+\cdots + \beta \left(\frac{n\sigma}{\mu+n\sigma}\right)^n\left(\frac{(m\gamma)^{m-1}}{(\mu+\alpha+m\gamma)^m}\right) & \cdots & 0  +\cdots + \beta\left(\frac{(m\gamma)^{m-2}}{(\mu+\alpha+m\gamma)^{m-1}}\right) & \cdots & 0 \\
  0  & \cdots & 0 & \cdots &  0 \\
  \vdots & \vdots & \vdots & \ddots & \vdots  \\
  0 & \cdots  & 0 &  \cdots & 0 \\
 \end{pmatrix}. 
\]
}
Therefore, the only nonzero eigenvalue is
\[
\beta \left(\frac{n\sigma}{\mu+n\sigma}\right)^n\left(\frac{1}{\mu+\alpha+m\gamma}\right) \left[1 + \frac{m\gamma}{\mu+\alpha+m\gamma} + \left(\frac{m\gamma}{\mu+\alpha+m\gamma}\right)^2 + ... + \left(\frac{m\gamma}{\mu+\alpha+m\gamma}\right)^{m-1} \right].
\]

 Summing the geometric series, the eigenvalue simplifies to 
\[
 \frac{\beta}{\mu+\alpha}\left(\frac{n\sigma}{\mu+n\sigma}\right)^n\left(1-\left(\frac{m\gamma}{\mu+\alpha+m\gamma}\right)^m\right),
 \]
our final $R_0$ value.

\section{Infectious Steady State with Applying Vaccine on Susceptible Class}
\label{VonS}

The infectious steady state for the multi-infected compartment model with a vaccine applied to the susceptible class (see Fig.~\ref{fig:compflowV}) is 
\begin{equation}
\sum_{i = 1}^n I_i^* = \left(\frac{n\sigma}{\mu+n\sigma}\right)^n\left(1-\left(\frac{m\gamma}{\mu+m\gamma}\right)^m\right)-\frac{\mu+\nu}{\beta}.\\
\end{equation}

The delay system's infectious steady state is
\begin{equation}
I^* = e^{-(\mu +\alpha)\tau_1}\left(1 - e^{-\mu\tau_2}\right) - \frac{\mu}{\beta}.\\
\end{equation}

\section{Infectious Steady States with Applying Quarantine on Exposed Class}
\label{SSQonE}
The infectious steady state for the multi-infected compartment model when a control measure, quarantine, is applied to the exposed class rather than the infectious class (see Fig.~\ref{fig:compflowQonE}) is
\begin{equation}
\sum_{i = 1}^m I_i^* = \left(\frac{n\sigma}{\mu + \alpha + n\sigma}\right)^n\left[1 - \left(\frac{m\gamma}{\mu + m\gamma}\right)^m\right] - \frac{\mu}{\beta}.
\end{equation}

The delay system's infectious steady state is
\begin{equation}
I^* = e^{-(\mu +\alpha)\tau_1}\left(1 - e^{-\mu\tau_2}\right) - \frac{\mu}{\beta}.\\
\end{equation}

%
%

\begin{acknowledgements}
AB and LBS were supported by NSF Grant No.~DMS-1715651.
\end{acknowledgements}

\bibliographystyle{spbasic}      





\end{document}